\documentclass[%
 reprint,
 showpacs,
 amsmath,amssymb,
prb,
]{revtex4-1}

\usepackage{dcolumn}
\usepackage[dvipdfmx]{graphicx}
\usepackage{mathrsfs}
\usepackage{accents}
\usepackage{bm}
\usepackage{color}

\makeatletter
\def\btt#1{\texttt{\@backslashchar#1}}
\DeclareRobustCommand\bblash{\btt{\@backslashchar}} \makeatother

\begin{document}

\title{Effects of Surface Roughness on Paramagnetic Response of Small Unconventional Superconductors}

\author{Shu-Ichiro Suzuki$^1$ 
and Yasuhiro Asano$^{1,2,3}$}
\affiliation{$^1$Department of Applied Physics,
Hokkaido University, Sapporo 060-8628, Japan}
\affiliation{$^2$Center for Topological Science \& Technology,
Hokkaido University, Sapporo 060-8628, Japan}
\affiliation{$^3$Moscow Institute of Physics and Technology, 
141700 Dolgoprudny, Russia}
\date{\today}

\begin{abstract}
We theoretically study effects of surface roughness on the magnetic response of small unconventional superconductors by 
solving the Eilenberger equation for the quassiclassical Green function and the Maxwell equation for the vector potential 
simultaneously and self-consistently. The paramagnetic phase of spin-singlet $d$\,-wave superconducting disks is drastically 
suppressed by the surface roughness, whereas that of spin-triplet $p$\,-wave disks is robust even in the presence of the 
roughness.  Such difference derives from the orbital symmetry of paramagnetic odd-frequency Cooper pairs appearing at the 
surface of disks. The orbital part of the paramagnetic pairing correlation is $p$\,-wave symmetry in the $d$\,-wave disks, 
whereas it is $s$\,-wave symmetry in the $p$\,-wave ones. Calculating the free-energy, we also confirm that the paramagnetic 
state is more stable than the normal state, which indicates a possibility of detecting the paramagnetic effect in experiments. 
Indeed our results are consistent with an experimental finding on high-$T_c$ thin films.  
\end{abstract}


\maketitle

\section{Introduction}
Diamagnetic response to an external magnetic field is a fundamental property of all superconductors~\cite{tinkham}. 
The Meissner current (coherent motion of the Cooper pairs) screens a weak magnetic field at a surface of a superconductor. 
As a result, the phase coherence of superconducting condensate is well preserved far away from the surface. A number of 
experiments, however, have reported the paramagnetic response of small superconductors and mesoscopic proximity structures~
\cite{W.Braunisch_PRL_1992,B.Schliepe_PRB_1993,D.J.Thompson_PRL_1995,A.K.Geim_Nature_1998,P.Visani_PRL_1990,A.C.Mota_Physica_1994}.

Recent theoretical studies have suggested the existence of paramagnetic Cooper pairs in inhomogeneous
superconductors~\cite{Y.Asano_PRL_2011,S.Mironov_PRL_2012,yoko1,melnikov12,Y.Asano_PRB_2012,asano14}. A spatial gradient 
of the superconducting order parameter induces subdominant pairing correlations. The pairing symmetry of such induced 
Cooper pairs is different from that of principal Cooper pairs in bulk superconducting state 
~\cite{Y.Tanaka_PRL_2007,Y.Tanaka_PRB_2007,asano14}. For example, the principal Cooper pairs in high-$T_c$ superconductors 
belong to the spin-singlet $d$\,-wave (even parity) class. In (110) direction of high-$T_c$ cuprate, a surface acts as 
a pair breaker and suppresses the pair potential drastically. Simultaneously, the spin-singlet odd-parity pairs are 
locally induced at the surface as a subdominant correlation. A surface generates odd-parity pairing correlations from 
the $d$\,-wave even-parity correlation because the surface breaks inversion symmetry locally. Since the pairing correlation 
function must be antisymmetric under the permutation of two electrons, the induced pairs have the odd-frequency symmetry~
\cite{berezinskii}. To our knowledge, such induced odd-frequency pairs indicate the paramagnetic response to an external 
magnetic field. Odd-frequency Cooper pairs can be generated also from conventional superconductors in the presence of 
spin-dependent potentials~\cite{bergeret}. 

In a previous paper~\cite{S.I.Suzuki_PRB_2014}, we have shown that magnetic susceptibility of small enough unconventional 
superconducting disks can be paramagnetic at a sufficiently low temperature. Odd-frequency Cooper pairs induced by a 
surface are responsible for the unusual paramagnetic Meissner effect. The magnetic response of Cooper pair is well 
characterized by so called ``pair density" which is defined by diagonal elements of the response function to a magnetic 
field. Even-frequency Cooper pairs have a positive pair density, whereas induced odd-frequency pairs have a negative pair 
density. So far an experiment has reported the decrease of pair density at low temperature in high-$T_c$ superconducting 
films on which internal surfaces are introduced by the heavy-ion irradiation~\cite{H.Walter_PRL_1998}. Thus our theoretical 
results are consistent with the experiment at least qualitatively. However, signs of the paramagnetic effect in the experiment~
\cite{H.Walter_PRL_1998} is much weaker than our theoretical prediction. The discrepancy may come from sample quality at 
surfaces. Artificially introduced internal surfaces can be very rough in the experiment, whereas surfaces are specular 
in the theory. Actually, several theories have pointed out that the surface roughness affects properties of the surface 
Andreev bound states of a high-$T_c$ superconductor~
\cite{A.Poenicke_PRB_1999,Yu.S.Barash_PRB_2000,Y.Tanaka_PRB_2001,A.Zare_PRB_2008,BS}.

The purpose of this paper is to clarify effects of surface roughness on the paramagnetic Meissner response of small 
unconventional superconductors. We consider a two-dimensional superconducting disk with spin-singlet $d$\,-wave symmetry 
or spin-triplet $p$\,-wave one. In numerical simulation, we solve the Eilenberger equation and the Maxwell equation 
simultaneously and self-consistently. Surface roughness is considered through an impurity self-energy within the 
self-consistent Born approximation. We find that the surface roughness suppresses drastically the paramagnetic response 
of a spin-singlet $d$\,-wave superconducting disk. On the other hand in a spin-triplet $p$\,-wave disk, the paramagnetic 
property is robust even in the presence of surface roughness. The induced odd-frequency pairing correlation has $p$\,-wave 
symmetry in the former, whereas it has $s$\,-wave symmetry in the latter. In addition, we also confirm that the paramagnetic 
superconducting states are more stable than the normal state by calculating free-energies.

This paper is organized as follows.    
In Sec.~II, we explain the theoretical method to analyze the magnetic response of small superconducting disks. 
In Sec.~III, we discuss the magnetic response of small superconducting disks with a rough surface.
In Sec.~IV, we discuss the stability of a paramagnetic state by calculating 
the free-energies of a superconducting state. We summarize this paper in Sec.~V. 

\section{Formulation}
\begin{figure}[tb]
  \begin{center}
    \includegraphics[width=0.45\textwidth]{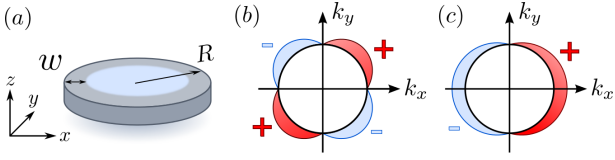}
    \caption{ 
			(a) Schematic figure of a superconducting disk with rough surface, with $R$ and $w$ are 
			the radius of the disk and the width of the disordered region, respectively. 
			An external magnetic field is applied in the $z$ direction. 
			The origin of a spatial coordinate is located at the center of the disk. 
      The pair potential in momentum space for the $d$\,-wave superconductor and 
			that for $p$\,-wave one are illustrated in (b) and (c), respectively.}
     \label{fig:schematic}
  \end{center}
\end{figure}

Let us consider a superconducting disk in two-dimension as shown in Fig.~\ref{fig:schematic}(a), 
where $R$ is the radius of the disk. 
To describe rough surfaces, we introduce random impurity potentials near the surface.
The width of the disordered region $w$ is measured from the surface as shown in Fig.~\ref{fig:schematic}(a). 
An external magnetic field $H^\mathrm{ext}$ is applied in the
$z$ direction. Throughout this paper, we use a unit of $\hbar =c=k_B=1$ with $k_B$ and $c$ 
being the Boltzmann constant and the speed of light, respectively. 

\begin{figure*}[t]
  \begin{center}
    \includegraphics[width=0.9\textwidth]{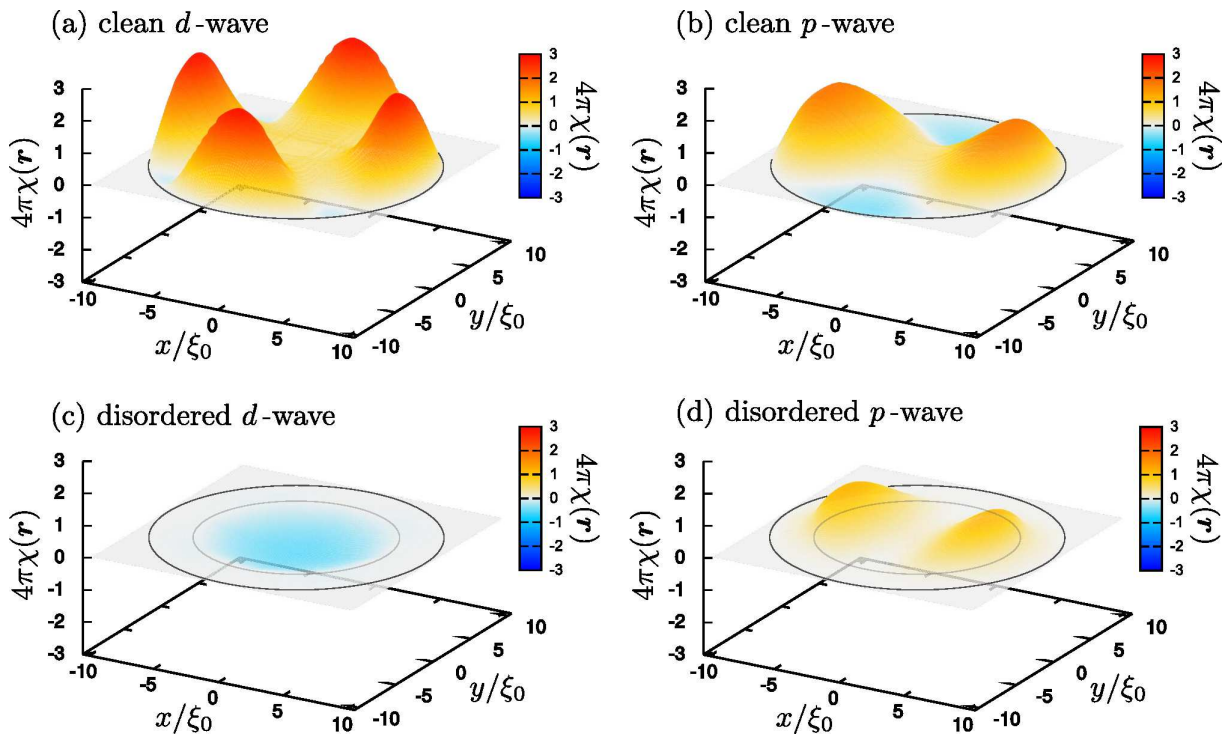}
      \caption{ 
				Local susceptibilities of the small superconducting disks. 
				The results of a $d$\,-wave and those of a $p$\,-wave superconductor with a clean surface 
				(i.e., $w=0$, $\xi/\ell = 0$) are presented in (a) and (b), respectively.
				The results of a $d$\,-wave and those of a $p$\,-wave superconductor 
				with rough surface ($w=3\xi_0$, $\xi_0/\ell = 1.0$) are demonstrated in (c) and (d), respectively. 
				The parameters used in the simulation are $R=10\xi_0$, $\lambda_L=5\xi_0$, $\omega_c=10\Delta_0$, 
				$H^\mathrm{ext}= 0.01H_{c_1}$ and $T=0.1T_c$. }
      \label{fig:chi_r}
  \end{center}
\end{figure*}

Superconducting states in equilibrium are described by solutions of the quasiclassical 
Eilenberger equation~\cite{Eilen}, 
\begin{align}
  & i v_F\bm{k} \cdot 
  \bm{\nabla}_{\bm{r}} \, \check{{g}}
  +\left[ \check{H} + \check{\Sigma} \, , \check{g} \, \right]=0,
  \label{eilenberger_eq}
\end{align}
where $v_F$ is the Fermi velocity, $\bm{k}$ is the unit vector on the Fermi surface. 
$\check{g}$ and $\check{H}$ are defined as follows, 
\begin{align}
  \check{g}(\bm{r},{\bm{k}},i\omega_n)
  &=\left[\begin{array}{cc}
                 \hat{g} (\bm{r},{\bm{k}},i\omega_n) & 
                 \hat{f} (\bm{r},{\bm{k}},i\omega_n) \\
    -\undertilde{\hat{f}}(\bm{r},{\bm{k}},i\omega_n) & 
    -\undertilde{\hat{g}}(\bm{r},{\bm{k}},i\omega_n)
  \end{array}\right],
  \label{g_def} \\[2mm]
  \check{H}(\bm{r},{\bm{k}},i\omega_n)
  &=\left[\begin{array}{cc}
                \hat{\xi   } (\bm{r},{\bm{k}},i\omega_n) & 
                \hat{\Delta} (\bm{r},{\bm{k}})           \\
    \undertilde{\hat{\Delta}}(\bm{r},{\bm{k}})           & 
    \undertilde{\hat{\xi   }}(\bm{r},{\bm{k}},i\omega_n)
  \end{array}\right],
  \label{H_def}  \\[2mm]
  \hat{\xi}(\bm{r},{\bm{k}},i\omega_n)
  & = \left( i\omega_n + ev_F \bm{k} \cdot \bm{A}(\bm{r}) \right) \hat{\sigma}_0 ,
	\label{shift}
\end{align}
where $ \omega_n = (2n+1)\pi T $ is the fermionic Matsubara frequency, 
$n$ is an integer number, and $T$ is a temperature. 
In this paper, the symbol $\check{\cdots}$ represents a $4 \times 4$ matrix structure, 
$\hat{\cdots}$ represents a $2 \times 2$ matrix structure in spin space and 
$ \hat{\sigma}_0$ is the identity matrix in spin space. 
A vector potential is denoted by $ \bm{A}(\bm{r})$. 
We introduced a definition $\undertilde{K}(\bm{r},\bm{k},i\omega_n) \equiv 
K^\ast(\bm{r},-{\bm{k}},i\omega_n)$ for all functions $K(\bm{r},\bm{k},i\omega_n)$. 
Effects of rough surfaces are taken into account through an impurity 
self-energy of a quasiparticle defined by, 
\begin{align}
  \check{\Sigma} (\bm{r}, i \omega_n) = 
    \Theta \left( |\bm{r}|-R+w \right) ~
    \frac{i }{2 \tau_0}
    \int \frac{d \bm{k}}{2\pi}~
    \check{g} (\bm{r}, \bm{k}, i\omega_n), 
\end{align}
where $\tau_0$ is the life time of a quasiparticle 
and $\Theta( x )$ is 
the Heviside step function. 
The mean free path of a quasiparticle is defined by $ \ell = v_F \tau_0 $ 
in the disordered region.
%
The anomalous Green function $\hat{f}(\bm{r},\bm{k}, i\omega_n)$ is 
originally defined by an average of two annihilation operators of an electron. 
The relation 
\begin{align}
  \hat{f}(\bm{r},\bm{k}, i\omega_n) = 
  -\hat{f}^\mathrm{\,T}(\bm{r}, -\bm{k}, -i\omega_n),\label{fd}
\end{align}
represents the antisymmetric property of the anomalous Green function 
under the permutation of two electrons,
where $\mathrm{T}$ represents the transpose of a matrix. 

The direction of $\boldsymbol{k}$ in two-dimensional momentum space is 
represented by an angle $\theta$ measured from the $x$ axis, (i.e., 
$k_x = \cos \theta$ and $k_y = \sin \theta$). 
In what follows, we consider two unconventional superconductors with 
different pairing symmetries.
One is spin-singlet $d$\,-wave symmetry 
$\hat{\Delta}(\bm{r},\theta) = \Delta(\bm{r}) \sin(2\theta) \, i \hat{\sigma}_2$.
The other is spin-triplet $p$\,-wave symmetry 
$\hat{\Delta}(\bm{r},\theta) = \Delta(\bm{r}) \cos(\theta) \, \hat{\sigma}_1$, 
where $ \hat{\sigma}_j$ for $j = 1$-$3$ are the Pauli matrices in spin space.  
A $d$\,-wave and a $p$\,-wave pair potentials in momentum space are shown schematically in 
Fig.~\ref{fig:schematic}(b) and (c), respectively. 
We do not consider any spin-dependent potentials in this paper.
The matrix structure of Green function is represented by
\begin{align}
  \hat{g} (\bm{r},\theta,i\omega_n) 
	=& g (\bm{r},\theta,i\omega_n) \hat{\sigma}_0,	\\
  \hat{f} (\bm{r},\theta,i\omega_n) 
	=& f (\bm{r},\theta,i\omega_n) \times
  \left\{ \begin{array}{cc} 
       \hat{\sigma}_2 & \text{for spin-singlet} \\
    -i \hat{\sigma}_1 & \text{for spin-triplet}
  \end{array}\right. ,
\end{align}
with scalar Green functions $g (\bm{r},\theta,i\omega_n)$ and $f (\bm{r},\theta,i\omega_n)$. 
Spatial dependence of $\Delta(\bm{r})$ is determined self-consistently from the gap equation 
\begin{gather}
  \Delta ( \bm{r} )
  = \pi N_0 g_0 T \sum_{\omega_n } \int \frac{ d \theta }{2 \pi }
  f ( \bm{r}, \theta, i\omega_n ) V_x(\theta),
  \label{eq:gap}
\end{gather}
where $N_0$ is the density of states per spin of normal metal at the Fermi level, 
$g_0$ is the coupling constant, and $V_x$ represents attractive electron-electron interactions 
with $x = p$\,-wave or $d$\,-wave indicating pairing symmetries. 
The interaction kernel $V_x$ depends upon pairing symmetries as 
\begin{align}
  V_x(\theta) = \left\{ 
  \begin{array}{cl}
    2 \cos   \theta  & \text{for ~$x=p$\,-wave} \\
    2 \sin (2\theta) & \text{for ~$x=d$\,-wave} 
  \end{array}. \right.
\end{align}
The constant $N_0 g_0$ is determined by 
\begin{align}
  (N_0 g)^{-1} =
  \ln \left( \frac{T}{T_c} \right) + 
  \sum_{0 \le n < \omega_c / 2 \pi T } 
  \frac{1}{n+1/2} ~, 
\end{align}
with $T_c$ and $\omega_c$ being the transition temperature and the cut-off energy, 
respectively. 

In a type-II superconductor, an electric current is represented by 
\begin{align}
 \bm{j}(\bm{r})=
 \frac{\pi e v_F N_0}{2i} T \sum_{\omega_n} \int \frac{ d \theta }{2 \pi }
 \textrm{Tr} \left[ \check{T}_3\, \bm{k}\, \check{g}( \bm{r}, \theta, \omega_n )\right],
 \label{eq:j}
\end{align}
with $\check{T}_3 = \mathrm{diag}[\hat{\sigma}_0,-\hat{\sigma}_0]$. 
From Eq.~(\ref{eq:j}) and the Maxwell equation $\bm{\nabla} \times \bm{H}(\bm{r}) = 
4 \pi \bm{j}(\bm{r}) $, we obtain spatial profiles of 
a vector potential $\bm{A} (\bm{r})$ and a local magnetic field $\bm{H}(\bm{r})$.  
A local magnetic susceptibility is defined by 
\begin{align}
   \chi (\bm{r})= \frac{1}{4 \pi}
  \frac{ H(\bm{r}) - H^\mathrm{ext} }{ H^\mathrm{ext} }, 
\end{align}
where $H^\mathrm{ext}$ is an amplitude of external magnetic field applied in the $z$ direction. 
By integrating the local susceptibility, we obtain a susceptibility of a whole disk 
$X$ as
\begin{align}
 X = 
  \frac{1}{\pi R^2}
  \int_{|\bm{r}| \le R} d\bm{r} \, 
  \chi (\bm{r}). 
\end{align}

To solve the Eilenberger equation Eq.~(\ref{eilenberger_eq}) in a disk geometry, 
we use a Riccati parametrization~\cite{Scho1,Scho2,M.Eschrig_PRB_2009} and a numerical method 
discussed in Ref.~\onlinecite{Y.Nagai_PRB_2012}. 
Using the parametrization, the Eilenberger equation can be separated into two Riccati type 
differential equations. When we solve the Riccati type equation along a long enough 
quasiclassical trajectory, solutions of the equation do not depend on initial conditions~\cite{Y.Nagai_PRB_2012}. 
In this paper, the length of classical trajectories is more than 30 times of the coherence length.
Solving the Eilenberger equation and the Maxwell equation, 
we determine the pair potential, the vector potential, and the self-energy self-consistently 
with one another. 
At surfaces, we consider specular classical trajectories 
for calculating the Green functions~\cite{Y.Nagai_PRB_2012}. 
The vector potential out side of the superconducting disk is $\boldsymbol{A}(\boldsymbol{r}) = 
(H^\mathrm{ext}/2 ) ( -y \hat{ \boldsymbol{x} } + x \hat{ \boldsymbol{y} } )$ which gives 
a uniform magnetic field in the $z$ direction, where  
$\boldsymbol{x}$ and $\boldsymbol{y}$ are the unit vector in the $x$ and $y$ direction, respectively.

\begin{figure}[t]
  \begin{center}
    \includegraphics[width=0.48\textwidth]{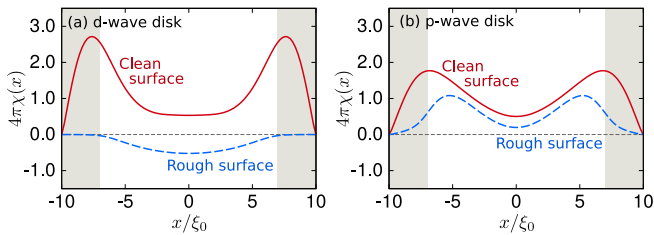}
    \caption{
			Local susceptibilities on the $x$ axis (i.e., $y=0$). The results for a $d$\,-wave disk 
			and for a $p$\,-wave disk are plotted in (a) and (b), respectively. 
			The all parameters are set to be the same as those of Fig.~\ref{fig:chi_r}. 
	  	The solid and the broken lines indicate the result for the disk with clean surface 
		  and those with the rough surface, respectively. 
	  	The shadowed areas indicate the disordered regions.}
    \label{fig:chi_x}
  \end{center}
\end{figure}
\begin{figure}[t]
  \begin{center}
    \includegraphics[width=0.48\textwidth]{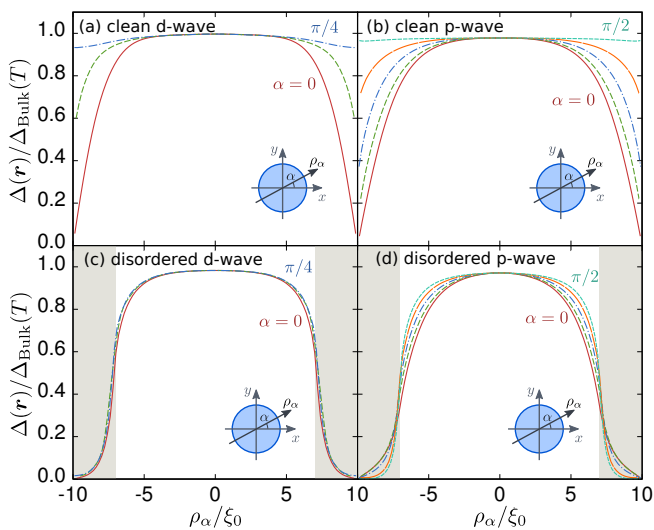}
    \caption{
			Amplitudes of pair potentials in real space for several azimuthal angle $\alpha$ measured 
			from the $x$ axis.
			The results of a $d$\,-wave disk with 
			$\ell/\xi_0 = 0.0$ and $\ell/\xi_0 = 1.0$ are shown in (a) and (c), respectively.
			Since the pair potential is four-fold symmetric, we plot its spatial profile  
			for $\alpha=0, \pi/8$, and $\pi/4$.
            The results of a $p$\,-wave disk with $\ell/\xi_0 = 0.0$ and $\ell/\xi_0 = 1.0$ are shown 
			in (b) and (d), respectively. 
			The pair potential for $\alpha = 0, \pi/8, \pi/4, 3\pi/8, \pi/2$ (from bottom to top) are shown under the 
			two-fold symmetry of results.  
			The pair potentials are normalized to $\Delta_\mathrm{Bulk}(T).$ 
			The all parameters are set to be the same as those of Fig.~\ref{fig:chi_r}. 
		The shadowed areas in (c) and (d) indicate the disordered regions.}
    \label{fig:del_r}
  \end{center}
\end{figure}
\section{Results}

Throughout this paper, we fix several parameters as 
$R=10\xi_0$ and $\omega_c=10\Delta_0$, where 
$\Delta_0$ is the amplitude of the pair potential at the zero temperature, 
and $ \xi_0 = \hbar v_F / 2 \pi T_c $ is the coherence length. 
Strength of the disorder is tuned by changing a parameter $\xi_0/\ell$.
Width of disordered region is $w=3 \xi_0$ because odd-frequency 
pairs induced by a surface localize within this range. 
All lengths are measured in units of $ \xi_0 $. 
The current density is normalized to $J_0 = \hbar c^2 / 4 \pi |e|\xi_0^3 $.
Here we express $\hbar$ and $c$ explicitly to avoid misunderstandings.
The characteristic length scale of the Maxwell equation is the penetration depth defined as 
$\lambda_L=(4\pi n e^2/mc^2)^{-1/2}$ and is a parameter in numerical simulations.
In this paper, we choose $\lambda_L=5\xi_0$ to realize type-II superconductors and fix
$H^\mathrm{ext}= 0.01H_{c_2}$.
Here $H_{c_2} = \hbar c/|e| \xi_0^2 $ is the second critical magnetic field. 
The first critical magnetic field at low temperature is estimated as $H_{c1} = H_{c_2}(\xi_0/\lambda_L)^2 \log(\lambda_L/\xi_0) 
\approx 0.03 H_{c2} > H^{\mathrm{ext}}$ at $\lambda_L=5\xi_0$. Thus vortices are not expected at low temperature.
We start numerical simulations with initial conditions of
a spatially uniform pair potential $\Delta (\boldsymbol{r}) = 
| \Delta_\mathrm{Bulk}(T) |$ and a homogeneous magnetic field $\boldsymbol{A}(\boldsymbol{r}) = 
(H^\mathrm{ext}/2 ) ( -y \hat{ \boldsymbol{x} } + x \hat{ \boldsymbol{y} } )$, where 
$\Delta_\mathrm{Bulk}(T)$ is the pair potential obtained in a homogeneous superconductor at a temperature $T$.  
If we choose an alternative initial condition hosting a vortex in a superconductor, a vortex state 
might be realized in numerical simulations~\cite{M.Ichioka_prb_1996} even for $H^{\mathrm{ext}}<H_{c1}$. 
Such vortex issue, however, goes beyond the scope of this paper. 

First we discuss calculated results  
for a small superconducting disk with a {\it specular} surface 
(i.e., $w=0$, $\xi_0 / \ell = 0$). 
Then we discuss effects of surface roughness by comparing 
numerical results in a disk with a rough surface with those with a specular one.

\begin{figure}[t]
  \begin{center}
    \includegraphics[width=0.46\textwidth]{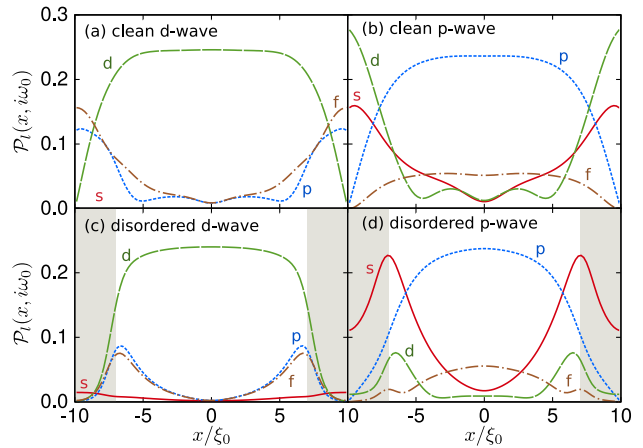}
    \caption{
			Spatial dependencies of the decomposed pairing functions at the lowest Matsubara frequency 
			on the $x$-axis for (a) a clean $d$\,-wave, (b) a clean $p$\,-wave, 
			(c) a disordered $d$\,-wave, and (d) a disordered $p$\,-wave disk. 
			The parameters are set to be the same values as those in Fig.~\ref{fig:chi_r}. 
		  The shadowed areas indicate the disordered regions. }
    \label{fig:pair_x}
  \end{center}
\end{figure}

\subsection{Disks with a specular surface}
\begin{figure}[t!]
  \begin{center}
  \includegraphics[width=0.48\textwidth]{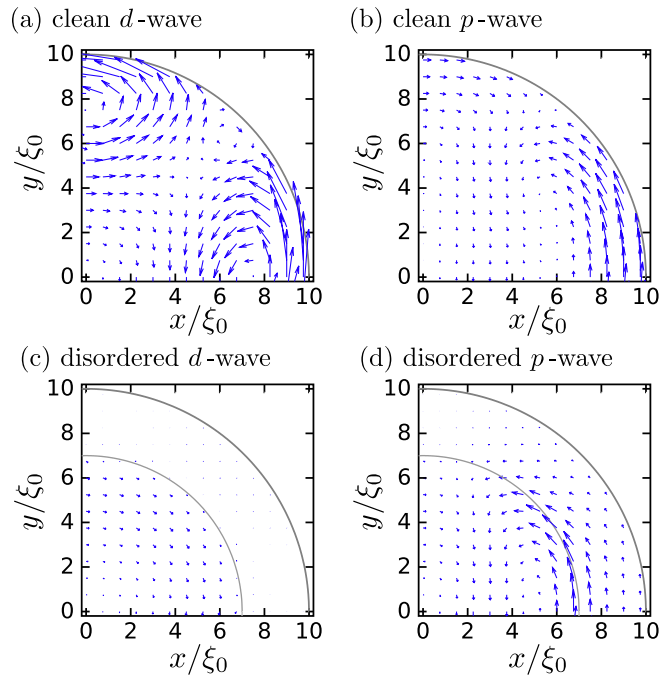}
	\caption{
		Spatial distribution of current density on a superconducting disk. The results of a $d$\,-wave and a $p$\,-wave 
		superconductor with clean surface ($w=0$, $\xi/\ell = 0.0$) are shown in (a) and (b), 
		respectively. The results with rough surface are shown in (c) and (d). 
		The outer circles indicate the edge of the disk and the inner circles in (c) and (d) 
		indicate the effective interface between the clean region and the disordered region. 
		The parameters are set to be the same values as those in Fig.~\ref{fig:chi_r}.}
  \label{fig:cur2}
  \end{center}
\end{figure}

We briefly explain the paramagnetic Meissner effect in a superconducting disk with a 
specular surface. 
Figure~\ref{fig:chi_r}(a) shows the local susceptibility in a $d$\,-wave superconductor. 
The result are four-fold symmetric reflecting the $d$\,-wave pair potential. The magnetic 
susceptibility is positive (paramagnetic) near four surfaces in the $x$ and the $y$ 
directions. In Fig.~\ref{fig:chi_x}(a), we show a spatial profile of 
the local susceptibility of Fig.~\ref{fig:chi_r}(a) along the $x$ axis at $y=0$.
We also show the amplitudes of pair potential in real space 
$| \Delta( \boldsymbol{r} ) |$ in Fig.~\ref{fig:del_r}(a). 
The pair potential is calculated along a trajectory $\rho_\alpha$ 
oriented by an angle $\alpha$ measured from the $x$ axis as shown in Fig.~\ref{fig:del_r}.
In a $d$-wave disk, the pair potential is four-fold symmetric.
The pair potential is strongly suppressed at the four surfaces in $x$ and $y$ directions  
as a result of appearing surface Andreev bound states~\cite{abs}, whereas it is totally 
constant in the directions in which the surface Andreev bound states are absent.
In contrast to the pair potential around a vortex core~\cite{M.Ichioka_prb_1996}, 
the results in Fig.~\ref{fig:del_r}(a) are anisotropic in real space.
As shown in Ref.~\onlinecite{asano14},
the spatial gradient of pair potential generates paramagnetic pairing correlations with odd-frequency symmetry.
Thus odd-frequency pairs are expected at four surfaces in the $x$ and $y$ directions, 
which explains inhomogeneous and angular anisotropic paramagnetic response in Fig.~\ref{fig:chi_r}(a).
In this paper, we analyze the frequency symmetries of Cooper pairs by 
decomposing pairing functions into a series of Fourier components. 
In a $d$\,-wave disk, the anomalous Green function are described by two components
\begin{align}
\hat{f}(\bm{r},\theta, i\omega_n)= \left[f_{\mathrm{ep}}(\boldsymbol{r},\theta,i\omega_n) 
+ f_{\mathrm{op}}(\boldsymbol{r},\theta,i\omega_n)\right] \hat{\sigma}_2, \label{feo}
\end{align}
where $f_{\mathrm{ep}}$ is an even-parity ($d$\,-wave) function representing the principal 
pairing correlation and $f_{\mathrm{op}}$ is an odd-parity function representing 
the induced pairing component at a surface.  To satisfy Eq.~(\ref{fd}), 
$f_{\mathrm{op}}$ must be an odd function of $\omega_n$. 
We decompose pair functions $f( \boldsymbol{r}, \theta, i\omega_n )$ as 
\begin{align}
	&\mathcal{P}_l
	( \boldsymbol{r}, i\omega_n )
	= 2 \sqrt{ C_l^2 + S_l^2 } , 
	\\
	&S_l = \int \frac{d \theta}{2 \pi}~
         \mathrm{Re} \left[ 
        		f( \boldsymbol{r}, \theta, i\omega_n )
         \right]
				\sin \left( l \theta \right),
	\\
	&C_l = \int \frac{d \theta}{2 \pi}~
         \mathrm{Re} \left[ 
        		f( \boldsymbol{r}, \theta, i\omega_n )
         \right]
				\cos \left( l \theta \right),
\end{align}
where $ l = 0, 1, 2 $ and 3 correspond to $s$\,-, $p$\,-, $d$\,- and $f$\,-wave
orbital functions, respectively. 
In the presence of a magnetic field, the imaginary part of $f (\boldsymbol{r}, \theta, i\omega_n)$ 
is induced by the vector potential as analytically shown in Appendix.
We focus only on the real part of $f$ to analyze pairing symmetries.
Figure~\ref{fig:pair_x}(a) indicate the spatial profile of $\mathcal{P}_l (x, i \omega_0)$ 
at the lowest Matsubara frequency as a function of $x$ at $y=0$. 
The pairing functions of induced Cooper pairs
have $p$\,- and $f$\,-wave symmetries and their amplitudes localize near the surface. 
Such odd-frequency Cooper pairs shows the paramagnetic response to a magnetic field. 
The surface also generates spin-singlet $s$-wave correlation. Its amplitude, however, is 
too small to confirm at a scale of plot in Fig.~\ref{fig:pair_x}(a). 
Figure~\ref{fig:cur2}(a) shows the spatial distribution of electric current on 
a $d$\,-wave disk. The diamagnetic current flows at the central region  
because of the usual Meissner effect. Near the surfaces in the $x$ and $y$ directions, 
however, the current flows the opposite direction to the Meissner current. 
Therefore a small $d$\,-wave superconductor can be paramagnetic due to 
the induced odd-frequency Cooper pairs at its surface.

A spin-triplet $p$\,-wave disk also indicates the similar paramagnetic effect 
as shown the results of magnetic susceptibility in Figs.~\ref{fig:chi_r}(b), 
its spatial profile on the $x$ axis in \ref{fig:chi_x}(b), pair potential in \ref{fig:del_r}(b) 
and electric current in \ref{fig:cur2}(b). 
The results are two-fold symmetric reflecting the $p$\,-wave 
superconducting pair potential. The paramagnetic effect can be seen near the surface 
in the $x$ direction because of induced odd-frequency Cooper pairs. 
The anomalous Green function is represented by Eq.~(\ref{feo}) with replacing 
$\hat{\sigma}_2$ by $-i\hat{\sigma}_1$. In the $p$\,-wave case, $f_{\mathrm{ep}}$ represents 
induced pairing correlations and is an odd function of $\omega_n$. 
As shown in Fig.~\ref{fig:pair_x}(b), $f_{\mathrm{ep}}$ mainly consists of 
$s$\,- and $d$\,-wave pairing correlations. 

In the end of this subsection, we summarize an important difference 
between the paramagnetic effect of a $d$\,-wave superconductor and that of a $p$\,-wave one.
In a $d$\,-wave disk, surface odd-frequency Cooper pairs
have a $p$\,- or $f$-wave symmetry~\cite{Y.Tanaka_PRB_2007,S.I.Suzuki_PRB_2014}. 
In a $p$\,-wave disk, on the other hand, $s$\,- or $d$\,-wave odd-frequency 
Cooper pairs are responsible for the paramagnetic 
effect~\cite{Y.Tanaka_PRB_2007,S.I.Suzuki_PRB_2014}. 
In the next subsection, we will show that the paramagnetic response of a disk with rough surface 
depends sensitively on the orbital symmetry of induced odd-frequency pairs at the surface.

\subsection{Disks with a rough surface}
Next, we discuss effects of the surface roughness on the magnetic response 
of a small superconductor. 
The calculated results of the local susceptibility for a $d$\,-wave  
superconducting disk with rough surface are shown in Fig.~\ref{fig:chi_r}(c), 
where we choose $\xi_0 / \ell = 1.0$. 
Comparing Fig.~\ref{fig:chi_r}(a) with (c), 
we can find that the surface roughness completely suppresses
the paramagnetic response at the four surfaces in a $d$\,-wave disk.
The central region of the disk with the rough surface recovers the usual diamagnetic response.
Such effect is demonstrated more clearly in
a spatial profile of local susceptibilities at $y=0$ in Fig.~\ref{fig:chi_x}(a), 
where the shadowed area indicates the disordered region. 
The amplitude of pair potential in real space is shown in Fig.~\ref{fig:del_r}(c).
The pair potential in the disordered region is totally suppressed because the random 
impurity potential acts as a pair breaker for $d$-wave Cooper pairs. 
Spatial profiles of decomposed pairing functions are shown in Fig.~\ref{fig:pair_x}(c). 
In the disordered region, a $d$\,-wave pairing function $\mathcal{P}_d$ is drastically 
suppressed due to impurity scatterings. The disordered region 
can be considered as a diffusive normal metal because the spatial profile
of order parameter is proportional to $\mathcal{P}_d$. 
Odd-frequency Cooper pairs are also fragile in the presence of surface roughness 
because they have $p$\,- or $f$\,-wave pairing symmetry. 
Therefore, both the paramagnetic current and the diamagnetic one disappear 
in the disordered region as shown in Fig.~\ref{fig:cur2}(c).
The magnetic property of a disk is determined by that at the clean central region 
where even-frequency $d$\,-wave Cooper pairs stay and contribute to the diamagnetic response.
We conclude that the paramagnetic effect in a $d$\,-wave disk is fragile in the presence 
of surface roughness because odd-frequency pairs have a $p$\,-wave 
or a $f$\,-wave orbital symmetry.

A $p$\,-wave disk indicates qualitatively different magnetic response from a $d$\,-wave one. 
The local susceptibility of a $p$\,-wave disk with rough surface is shown in 
Fig.~\ref{fig:chi_r}(d) and Fig.~\ref{fig:chi_x}(b) with a broken line. 
Although the surface is rough enough, 
a $p$\,-wave superconducting disk still show the strong paramagnetic response. 
The peak of the $\chi$ in Fig.~\ref{fig:chi_x}(b) 
shifts to inside of the disk in the presence of surface roughness. 
This suggests that the Andreev bound states appear at a boundary between 
the clean central region and the disordered surface region. 
Such Andreev bound states always accompany the paramagnetic odd-frequency 
Cooper pairs. In addition to this, the paramagnetic response 
in Fig.~\ref{fig:chi_x}(b) suggests 
the penetration of odd-frequency Cooper pairs into the surface disordered region.
The spatial profiles of the electric current in
Fig.~\ref{fig:cur2}(d) shows that the paramagnetic current flows not only in the clean region 
but also in the disordered one. 
Odd-frequency pairs in a $p$\,-wave disk survive even in the presence of surface roughness because 
they have $s$\,-wave orbital symmetry. Therefore, the paramagnetic effect in a $p$\,-wave disk 
is robust against the surface roughness.

\subsection{Temperature dependence}
\begin{figure}[t!] 
  \begin{center}
    \includegraphics[width=0.48\textwidth]{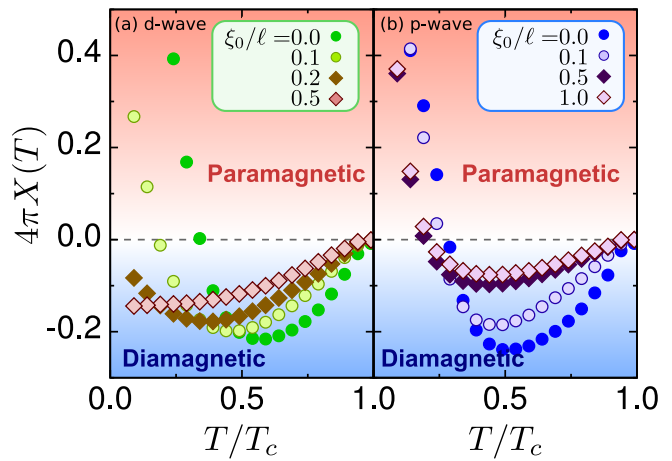}
  \caption{Temperature dependencies of the whole disk susceptibility $X(T)$ 
           for the $d$\,-wave disk (a) and the $p$\,-wave disk (b). }
  \label{fig:chi_tdep}
  \end{center}
\end{figure}

Here we discuss the magnetic susceptibility of a whole disk which is a measurable value in experiments.
The disk susceptibility in a $d$\,-wave superconductor and that in a $p$\,-wave one are plotted 
as a function of temperature in Fig.~\ref{fig:chi_tdep}(a) and (b), respectively. 
We present the results for several choices of the disorder $\xi_0/\ell$. 
In simulation, we first calculate the pair potential and the vector potential 
self-consistently at a temperature just below $T_c$ under an external magnetic 
field $H^\mathrm{ext}=0.01H_{c_2}$. 
Then the temperature is decreased with keeping $H^\mathrm{ext}$ unchanged. 
 In the clean limit ($\xi_0 / \ell = 0.0$), 
both a $d$\,-wave disk and a $p$\,-wave one show the usual diamagnetic response just below $T_c$. 
With decreasing the temperature, however, the sign of susceptibility changes 
around $T = T_p \sim 0.3T_c$ for both cases. 
Here, we define $T_p$ as a diamagnetic-paramagnetic crossover temperature. 
Blow $T_p$, superconducting disks show the anomalous paramagnetic response. 
The paramagnetic effect is stronger in lower temperature because  
 odd-frequency Cooper pairs energetically localize at the Fermi level.

In a $d$\,-wave disk, the reentrance is slightly suppressed in the presence of the 
moderate surface roughness with $\xi_0 / \ell = 0.1$ as shown in Fig.~\ref{fig:chi_tdep}(a).
When we increase the degree of the roughness further, the paramagnetic response gradually 
becomes weaker.
At $\xi_0 / \ell = 0.5$, the response is diamagnetic and the susceptibility recovers 
the monotonic temperature dependence which is usually observed in large enough superconductors 
in experiments.  In the $d$\,-wave superconductors, the specular Andreev reflection is necessary 
for forming the surface bound states at the zero-energy~\cite{asano02} and for appearing 
odd-frequency pairs. In other wards, odd-frequency pairs have $p$\,- or $f$\,-wave orbital 
symmetry.  Therefore the rough surface brakes odd-frequency pairs and suppresses 
the paramagnetic response. This conclusion is totally consistent with 
the experiment~\cite{H.Walter_PRL_1998}, where temperature dependences of the pair density 
on a high-$T_c$ superconducting film show a small reentrant behavior at low temperature. 
But the total pair density remains positive.  
Actually, the experimental data are very similar to the results for $\xi_0 / \ell = 0.2$ 
in Fig.~\ref{fig:chi_tdep}(a). 
The experimental results can be interpreted 
as an appearance of a small amount of odd-frequency pairs.  In the experiment, 
the surface roughness may partially breaks odd-frequency pairs because a number of the 
internal surfaces are introduced by the heavy ion irradiation.

On the contrary to a $d$\,-wave disks, the susceptibility of a $p$\,-wave disk $X(T)$ shows 
the reentrance and the crossover to the paramagnetic phase at low temperature 
for all $\xi_0/\ell$ as shown in Fig.~\ref{fig:chi_tdep}(b).
It has been pointed out that the surface Andreev bound states of a $p$\,-wave superconductor 
are robust under potential disorder because of the pure chiral property of 
surface bound states~\cite{ikegaya}. 
In other wards, odd-frequency pairs accompanied by the Andreev bound states 
have $s$\,-wave orbital symmetry~\cite{Y.Tanaka_PRB_2007,S.I.Suzuki_PRB_2014}. 
Since $s$\,-wave pairs is robust under the disordered potential, 
the paramagnetic effect of the $p$\,-wave superconductors persists even in the presence of 
surface roughness. 
We conclude that the robust paramagnetic response in a small size sample is 
a unique property of spin-triplet $p$\,-wave superconductors.
Such property would enable us to identify the spin-triplet $p$\,-wave superconductivity 
in experiments.

\section{Stability of Paramagnetic Superconducting States}

Generally speaking, a superconducting phase is more stable than a normal one 
as far as a superconductor is diamagnetic and homogeneous~\cite{tinkham}. 
Therefore a homogeneous paramagnetic superconducting phase is usually 
unstable. 
The calculated results in Sec.~III, however, show that the paramagnetic 
phase on a small superconducting disk is spatially inhomogeneous. 
In such situation, it would be worthy to check if 
the paramagnetic phase is a stable state at a free-energy minimum
or a metastable state corresponding to a free-energy local minimum. 
In this section, we discuss the stability of paramagnetic phase in small 
unconventional superconductors by calculating the free-energy in clean 
superconducting disks.

The free-energy is calculated from the quasiclassical Green functions~\cite{eilenberger2},  
\begin{align}
  &{F}_S- {F}_N 
	= \int d \boldsymbol{r} \;  \mathcal{F}(\boldsymbol{r}), 
	\\
  & \mathcal{F}(\boldsymbol{r}) 
	= \mathcal{F}_\Delta(\boldsymbol{r}) + \mathcal{F}_H(\boldsymbol{r}), 
	\\
  &\mathcal{F}_H(\boldsymbol{r}) 
	= \frac{\left\{ H(\boldsymbol{r}) - H^\mathrm{ext} \right\}^2}{8\pi}, 
	\label{eq:FE} 
	\\
  &\mathcal{F}_\Delta(\boldsymbol{r}) 
	= \mathcal{F}_f(\boldsymbol{r}) + \mathcal{F}_g(\boldsymbol{r}),
  \label{fdelta} 
	\\
  &\mathcal{F}_f(\boldsymbol{r}) = 
  \pi N_0  \int \frac{d\theta}{2\pi}\, 
  T\sum_{\omega_n} 
	\Delta^\ast{ (\bm{r},\theta)} 
	f(\bm{r},\theta,i\omega_n), \label{fdeltaf} 
	\\
  &\mathcal{F}_g(\boldsymbol{r}) = 
  4\pi N_0  \int \frac{d\theta}{2\pi}\, 
  T\sum_{\omega_n > 0}^{\omega_c} 
	\nonumber \\
	& \hspace{13mm}\times \int_{\omega_{n}}^{\omega_{c_2}} d\omega \;
    \textrm{Re} \left\{ 
    g (\bm{r},\theta,i\omega ) - 1 
    \right\}, 
		\label{fdeltag}
\end{align}
where $\mathcal{F}_\Delta (\bm{r}) $ is the condensation energy density of electron system 
and $\mathcal{F}_H (\bm{r}) $ is the energy density of a magnetic field. 
We introduce an additional energy cut-off 
$\omega_{c_2}$ to evaluate the integration in Eq.~(\ref{fdeltag}).
 In this paper, we set ${\omega}_{c_2} = 400 \Delta_0$ so that 
$\int d\bm{r} \mathcal{F} (\bm{r})$ reaches to a converged value. 
The free-energy densities normalized to $\mathcal{F}_0 = N_0 | \Delta_0 |^2 / 2$ 
which is the condensation energy density in a homogeneous $s$\,-wave superconductor. 
A temperature is set to be sufficiently low at $T=0.1T_c$ so that a superconductor 
is in the paramagnetic phase. 

\begin{figure}[t]
  \begin{center}
  \includegraphics[width=8cm]{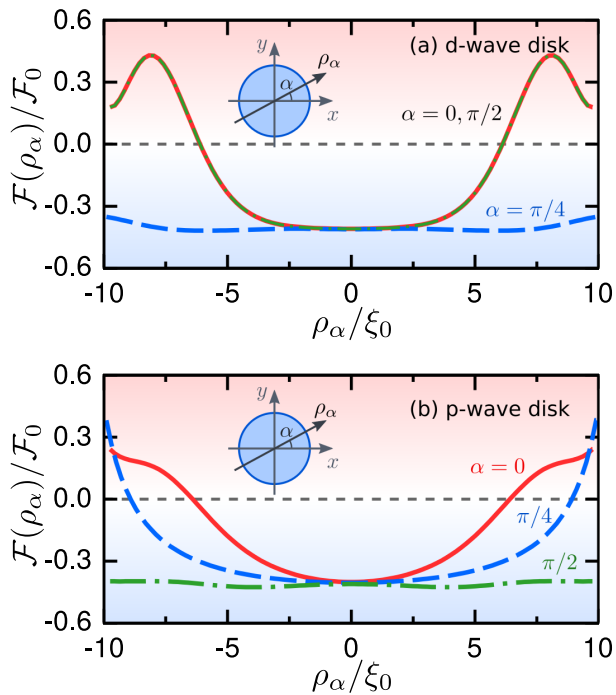}
  \caption{
Free-energy densities on a trajectory $\rho_\alpha$ of a $d$\,-wave disk (a) and 
a $p$\,-wave disk (b), where $\mathcal{F}(\bm{r})$ is normalized to 
  	$\mathcal{F}_0 = 	N_0 \Delta_0^2 /2$. The angle $\alpha$ is measured from the $x$ axis as shown in 
		schematics in the figures. 
		The temperature is set to $T=0.1T_c$ so that the superconducting disks show 
		the paramagnetic response. The second energy cut-off is set to $\omega_{c_2} = 400\Delta_0 $.
		The other parameters are fixed as $R=10\xi_0$, $\lambda_L=5\xi_0$, $\omega_c=10\Delta_0$, 
		$H^\mathrm{ext}= 0.01H_{c_1}$. }
  \label{fig:fe1}
  \end{center}
\end{figure}

The calculated results of the free-energy density for a $d$\,-wave disk and those for
a $p$\,-wave one are shown in Fig.~\ref{fig:fe1}(a) and (b), respectively. 
The free-energy density is calculated along a trajectory $\rho_\alpha$ 
oriented by an angle $\alpha$ which is measured from the $x$ axis as shown in Fig.~\ref{fig:fe1}.
In a $d$\,-wave disk, the results in Fig.~\ref{fig:fe1}(a) show that 
$\mathcal{F}(\boldsymbol{r})$ is negative around the disk center.
However it becomes positive near the surfaces at $\alpha=0$. 
On the other hand, $\mathcal{F}(\boldsymbol{r})$ at $\alpha=\pi/4$ 
is almost flat and is always negative along the trajectory because odd-frequency pairs are 
absent in this direction. 
The results for $\alpha=\pi/2$ are identical to those for $\alpha=0$ due to the four-fold symmetry. 
The free-energy density varies gradually from the line with $\alpha=0$ to that with 
$\alpha=\pi/4$ as increasing $\alpha$ from $\alpha=0$.
The free-energy of whole disk $\int d\boldsymbol{r} \mathcal{F}(\boldsymbol{r}) $ 
can be negative even though the disk is in the paramagnetic phase. 
In a $p$\,-wave disk, the increasing of $\mathcal{F}(\boldsymbol{r})$ occur only in two surfaces 
due to its $p$\,-wave symmetry. 
Therefore, we can conclude that both $d$\,- and $p$\,-wave superconducting states  
are more stable than a normal state even in their paramagnetic phase. 

\begin{figure}
  \begin{center}
  \includegraphics[width=8cm]{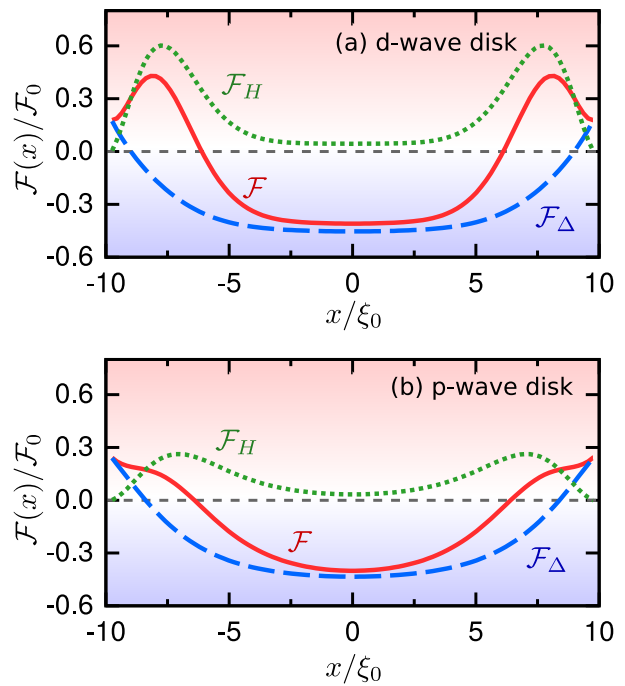}
  \caption{
		Condensation energy density and energy density of a magnetic field for $\alpha=0$. The results of 
		a $d$\,-wave disk and those of a $p$\,-wave disk are plotted in (a) and (b), respectively. 
		$\mathcal{F}$, $\mathcal{F}_\Delta$, and $\mathcal{F}_H$ are indicated in 
		solid lines, broken lines, and dotted lines, respectively. 
		All of the energy densities are normalized to $\mathcal{F}_0 = N_0 \Delta_0^2 /2$, 
		which is the condensation energy density of a homogeneous $s$\,-wave superconductor. 
		The all parameters are set to be the same as those of Fig.~\ref{fig:fe1}.}
  \label{fig:fe2}
  \end{center}
\end{figure}

To analyse the details of the free-energy density further,
we decompose the free-energy density at $\alpha=0$ into $\mathcal{F}_H$ 
and $\mathcal{F}_\Delta$ as shown in Fig.~\ref{fig:fe2}(a). 
When a superconductor shows perfect Meissner effect, 
the energy of a magnetic field becomes $\mathcal{F}_H = (H^\mathrm{ext})^2/8\pi$. 
In a small superconductor (i.e., $R \sim \lambda_L$), an external magnetic field 
penetrates into whole the disk. This suppresses $\mathcal{F}_H$ from $(H^\mathrm{ext})^2/8\pi$
at the center of the disk. Near the surface, on the other hand, paramagnetic 
odd-frequency Cooper pairs attract a magnetic field, which increases $\mathcal{F}_H$ 
locally. The appearance of odd-frequency pairs also increases $\mathcal{F}_\Delta$ as discussed
in Appendix.
As a result, $\mathcal{F}(\boldsymbol{r})$ becomes positive at the disk surface in both 
Fig.~\ref{fig:fe2}(a) and (b).
The condensation energy $\mathcal{F}_\Delta$ is negative at the disk center. 
While, near the surface, $\mathcal{F}_\Delta$ increases due to the suppression of the pair potential 
there in both Fig.~\ref{fig:fe2}(a) and (b).
In appendix, we analytically calculate the Green functions and the free-energy density 
near the surface of a semi-infinite $p_x$-wave superconductor.
The results in Eq.~(\ref{eq:ap-fe}) indicates that the free-energy density is positive 
due to the appearance of odd-frequency Cooper pairs. 
The numerical results in Fig.~\ref{fig:fe2}(a) and (b) show that the free-energy of whole disk 
$\int \, d\boldsymbol{r} \mathcal{F}(\boldsymbol{r}) $ remains negative 
because odd-frequency pairs are confined only near the surface.  
Therefore the paramagnetic superconducting state on $d$\,- and $p$\,-wave disks 
is more stable than the normal state.

\begin{figure}[t]
  \begin{center}
  \includegraphics[width=8cm]{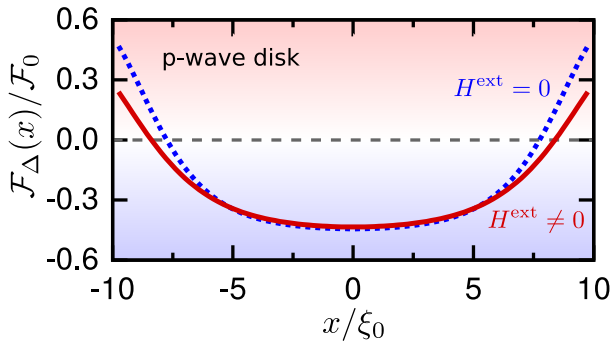}
	\caption{Condensation energy density $\mathcal{F}_\Delta$ obtained with a external field
  (solid line) and without a external field (dotted line). 
	Calculations are carried out for a $p$\,-wave superconducting disk. 
	The all parameters are set to be the same as those of Fig.~\ref{fig:fe1}.}
  \label{fig:fe3}
  \end{center}
\end{figure}

Next, we study energetic properties of odd-frequency pairs. 
In our simulation, it is possible to obtain two superconducting states: 
a superconducting state in the absence of a magnetic field $H^\mathrm{ext}=0$ 
and that in the presence of a magnetic field $H^\mathrm{ext}\neq 0$.
At $H^\mathrm{ext}=0$, there is no electric current everywhere. 
Superconducting states at $H^\mathrm{ext} \neq 0$, on the other hand, carry electric currents 
as shown in Fig.~5(a)-(d).
Here we compare condensation energies of such two different superconducting states 
as shown in Fig.~\ref{fig:fe3}, where we plot $\mathcal{F}_\Delta$ on a $p$\,-wave disk for 
$\alpha=0$. The solid line and the broken line indicate results for $H^\mathrm{ext} \neq 0$ 
and those for $H^\mathrm{ext} = 0$, respectively. 
As shown in Fig.~\ref{fig:fe3}, $\mathcal{F}_\Delta$ near surfaces 
for $H^\mathrm{ext} \neq 0$ is lower than that for $H^\mathrm{ext}=0$. 
Odd-frequency pairing state in the presence of electric currents 
is more stable than that in the absence of electric currents. 
This energetic property explains the paramagnetic property of odd-frequency Cooper pairs. 
The argument above is valid also for a $d$\,-wave disk.
In Appendix, we present analytical expression of difference 
between free-energy at $H^\mathrm{ext}=0$ and that at $H^\mathrm{ext}\neq 0$.
The results show that a magnetic field decreases the free-energy at low temperature
because odd-frequency Cooper pairs have the paramagnetic property.

\section{Conclusion}
We have theoretically studied effects of surface roughness on 
the anomalous paramagnetic response of small unconventional superconducting disks 
by using the quasiclassical Green function method. 
We conclude that the paramagnetic property of 
$p$\,-wave superconductors is robust under the surface roughness 
because the $p$\,-wave superconductors host the $s$\,-wave odd-frequency Cooper pairs at their surface.
On the other hand, the paramagnetic property in $d$\,-wave superconductor is fragile
in the presence of the surface roughness. In this case, the odd-frequency pairs 
at the surface have $p$\,-wave orbital symmetry. 
We have also confirmed that the paramagnetic superconducting phase are 
more stable than the normal state by calculating the free-energy. 
\begin{acknowledgments}
We are grateful to Y.~Tanaka, S.~Higashitani, Y.~Nagato, N.~Miyawaki, and S.~Ikegaya for helpful discussion. 
This work was partially supported by “Topological Quantum Phenomena” (Grant No. 22103002) 
Grant-in Aid for Scientific Research on Innovative Areas, KAKENHI (Grant No.  26287069) from 
the Ministry of Education, Culture, Sports, Science and Technology (MEXT) of Japan, and by the Ministry 
of Education and Science of the Russian Federation (Grant No. 14Y.26.31.0007). S.-I.~S. is supported 
in part by Grant-in-Aid for JSPS Fellows (Grant No. 15J00797) by Japan Society for the Promotion of Science (JSPS).
\end{acknowledgments}


\appendix*
\section{Analysis in a semi-infinite $p_x$-wave superconductor} 
In a semi-infinite superconductor in two-dimension, it is possible to obtain
analytic expression of Green functions in the clean limit. 
The evaluation of a electric current and free-energy by using analytical expressions would 
be helpful to understand numerical results in the text.

We assume that a superconductor occupies $x\geq 0$ and uniform in the $y$ direction.
An magnetic field applied in the $z$ direction and its vector potential is given by 
$\boldsymbol{A}= A(x) \hat{\boldsymbol{y}}$.
The Eilenberger equation in $2\times 2$ Nambu space reads,
\begin{align}
  & i v_F \boldsymbol{k} \cdot \nabla \hat{g} 
	  + [ \hat{H}, \hat{g}] =0,
		\\
  & \hat{g}(x,\boldsymbol{k},i\omega_n) 
    =\left[ \begin{array}{cc} 
				g                  & f \\
        s_p \undertilde{f} & -g 
		\end{array} \right]_{(x,\boldsymbol{k},i\omega_n)},
		\\
  & \hat{H}
    =\left[ \begin{array}{cc} 
				i\omega_n + ev_F \boldsymbol{k}\cdot \boldsymbol{A} & 
				i\Delta(x,\boldsymbol{k})                           \\
        is_p \undertilde{\Delta}(x,\boldsymbol{k})          & 
			  -i\omega_n - ev_F \boldsymbol{k}\cdot \boldsymbol{A} 
		\end{array} \right],\\
  & s_p=\left\{ \begin{array}{cc} 
		 1 & \textrm{even-parity} \\
    -1 & \textrm{odd-parity} 
	\end{array} \right. .
\end{align}
where a factor $s_p$ depends on the parity of order parameter and 
the Green functions satisfy $g^2+s_p \undertilde{f} f=1$.
The Green functions can be expanded with respect to the vector potential as
\begin{align}
  g=& g^{(0)} +  
	  (-i ev_F \boldsymbol{k}\cdot \boldsymbol{A}) 
		\partial_{\omega_n} g^{(0)} \nonumber \\
    & + \frac{1}{2} (-i ev_F \boldsymbol{k}\cdot \boldsymbol{A})^2 
		\partial^2_{\omega_n} g^{(0)}+ \cdots ,
		\label{ap-g}\\
  f=& f^{(0)} 
	  +(-i ev_F \boldsymbol{k}\cdot \boldsymbol{A}) 
	  \partial_{\omega_n} f^{(0)}  \nonumber\\
    & +\frac{1}{2} (-i ev_F \boldsymbol{k}\cdot \boldsymbol{A})^2 
		\partial^2_{\omega_n} f^{(0)} + \cdots,
		\label{ap-f}
\end{align}
because a vector potential shifts the Matsubara frequency,
where $g^{(0)}$ and $f^{(0)}$ are the Green function in the absence of a vector potential. 
In what follows, we omit ``(0)" from the Green function for simplicity.
We note in Eq.~(\ref{ap-f}) that parity and frequency symmetry of the second term 
on the right-hand side are opposite to those of the first term 
because $\boldsymbol{k}$ is a odd-parity function and $\partial_{\omega_n} f $ changes 
a frequency symmetry\cite{asano14}.
The imaginary part of an anomalous Green function 
represents a pairing correlation deformed by a vector potential.

In the case of a $p_x$-wave superconductor, 
it is possible to obtain a reasonable solution of the Eilenberger equation at $\boldsymbol{A}=0$. 
When we assume the spatial dependence of the pair potential as
\begin{align}
	\Delta(x,\theta)	
	=\Delta(\theta) \tanh \left( x / \xi \right), 
\end{align}
with $\xi = v_F/\Delta_0$, 
the Green functions are represented by~\cite{Scho2}
\begin{align}
  g(x,\theta,i\omega_n)= \,
	&\frac{\omega_n}{\Omega} 
	+\frac{\Delta^2(\theta)}{2\omega_n\Omega}
	\cosh^{-2}\left(\frac{x}{\xi}\right),
  \label{ap-gsh}\\
  f_{\textrm{P}}(x,\theta,i\omega_n)=\,
	&\frac{\Delta(\theta)}{\Omega} 
	\tanh\left(\frac{x}{\xi}\right),
	\label{ap-fp}\\
  f_{\textrm{I}}(x,\theta,i\omega_n)=
	& -\frac{\Delta^2(\theta)}{2\omega_n \Omega} 
	\cosh^{-2}\left(\frac{x}{\xi}\right),
  \label{ap-fi}
	%
\end{align}
where $\Omega=[ \, \omega_n^2 + \Delta^2(\theta)\,]^{1/2}$, and $\Delta(\theta)=\Delta_0 \cos(\theta)$. 
The Green function $f_{\textrm{P}}$ represents the principal pairing correlation in the bulk state,
whereas $f_{\textrm{I}}$ represents the pairing correlation induced by a surface at $x=0$.
They are calculated from the anomalous Green function as
\begin{align}
	f_P (x, \theta, i\omega_n) =
	\frac{1}{2} \left. ( f + s_p \undertilde{f} ) \right|_{(x, \theta, i\omega_n)}, 
	\\
	f_I (x, \theta, i\omega_n) =
	\frac{1}{2} \left. ( f - s_p \undertilde{f} ) \right|_{(x, \theta, i\omega_n)}.
\end{align}
At the deep inside of the supercondcutor (i.e., $x \gg \xi$), we obtain $f_P = f$ and $f_I=0$. 

\subsection{Current density}
>From an expression of electric current in Eq.~(\ref{eq:j}),
we define a linear response function $\mathcal{R}_{\mu,\nu}$ by
\begin{align}
  j_\mu(\boldsymbol{r}) =
	& -\frac{e^2}{m} \mathcal{R}_{\mu,\nu} A_\nu, 
	\label{ap-ja}\\[2mm]
  \frac{\mathcal{R}_{\mu,\nu}}{n_e} = \,
	& 4\pi T\sum_{\omega_n} \int \frac{d\theta}{2\pi} 
	k_\mu k_\nu \partial_{\omega_n}
  g(\boldsymbol{r}, \theta, i\omega_n) 
	\label{rmunu}.
\end{align}
with $\boldsymbol{k}=(\cos\theta, \sin\theta)$ and $n_e=v_F^2N_0 m$ being 
an electron density in two-dimension.
The diagonal elements of the response function $\mathcal{R}_{\mu,\mu}$ correspond to 
so called pair density. In the present situation, by substituting Eq.~(\ref{ap-gsh}) 
into Eq.~(\ref{rmunu}), we obtain
\begin{align}
  \frac{\mathcal{R}_{y,y}}{n_e} = \, 
	& 1 - \kappa_1 \frac{\Delta_0}{\omega_0} 
	\cosh^{-2} \left( \frac{x}{\xi} \right),
  \label{ap-ryy}\\
  \kappa_1=&
	\int\frac{d\theta}{2\pi} 
	\sin^2(\theta)|\cos(\theta)|
	=\frac{2}{3\pi},
\end{align}
where $\omega_0=\pi T$ is a low energy cut-off in the Matubara summation.
Using the normalization condition, the integrand in Eq.~(\ref{rmunu}) can be represented in an alternative way,
\begin{align}
  \partial_{\omega_n} g = 
	\left[ 
		-f_{\textrm{P}} \partial_{\omega_n} f_{\textrm{P}} 
    +f_{\textrm{I}} \partial_{\omega_n} f_{\textrm{I}} 
	\right] / g.  
	\label{ap-pomg}
\end{align}
It is possible to confirm that 
$-f_{\textrm{P}} \partial_{\omega_n} f_{\textrm{P}}/g$ corresponds to 
the first term in Eq.~(\ref{ap-ryy}), whereas $ f_{\textrm{I}} \partial_{\omega_n} f_{\textrm{I}}/g$
contribute to the second term. 
In this way, we can confirm that induced odd-frequency Cooper pairs indicate 
paramagnetic response to an external magnetic field.
Eq.~(\ref{ap-ryy}) suggests that the paramagnetic response is stronger in lower 
temperature.

\subsection{Free-energy density}
Substituting Eqs.~(\ref{ap-gsh})-(\ref{ap-fi}) into Eqs.~(\ref{fdelta})-(\ref{fdeltag}),
we find that a free-energy density at $\boldsymbol{A}=0$
\begin{align}
  \mathcal{F}_f= \,& 
	N_0 \kappa_2 \Delta_0^2 
	\log \left( \frac{2\omega_c}{\Delta_0} \right) 
  \tanh^2 \left(\frac{x}{\xi}\right),
	\\
  \mathcal{F}_g=& 
	-N_0 \kappa_2 \Delta_0^2 
	\log \left( \frac{2\omega_c}{\Delta_0} \right)
	\tanh^2\left(\frac{x}{\xi}\right)
  \nonumber\\ &
	+{N_0\Delta_0^2}\kappa_2
	\left[ \,
	   \cosh^{-2}\left(\frac{x}{\xi}\right) 
	  -\frac{1}{2} \,
	\right],\\
  \kappa_2=& 
	\int\frac{d\theta}{2\pi} \cos^2(\theta)
	=\frac{1}{2}.
\end{align}
As a result, we obtain 
\begin{align}
  \mathcal{F}_\Delta
	={N_0\Delta_0^2}\kappa_2
	\left[ \,
	  \cosh^{-2}\left(\frac{x}{\xi}\right) 
	  -\frac{1}{2} \,
	\right].
	\label{eq:ap-fe}
\end{align}
The free-energy density becomes positive at $x=0$ due to appearance of odd-frequency pairs.

Contribution of a magnetic field to the free-energy can be evaluated by 
applying the expansion in Eqs.~(\ref{ap-g})-(\ref{ap-f}) onto Eqs.~(\ref{ap-gsh})-(\ref{ap-fi}).
Within the second order expansion, we find that both the pair potential obtained by 
Eq.~(\ref{eq:gap}) and $\mathcal{F}_f$ remain unchanged.
The second order correction to $\mathcal{F}_g$ is given by
\begin{align}
  \mathcal{F}_g^{(2)}=
	& \frac{1}{2}N_0\left[ev_FA(x) \right]^2 
	\left[ 
    1-\kappa_1 \frac{\Delta_0}{\omega_0} 
		\cosh^{-2}\left(\frac{x}{\xi}\right) 
	\right].
\end{align}
The results indicate that $\mathcal{F}_g^{(2)}$ can be locally negative (paramagnetic)
at low enough temperature. This explains the decrease of the free-energy 
in a magnetic field shown in Fig.~\ref{fig:fe3}.
We also obtain the electric current in Eqs.~(\ref{ap-ja}) and (\ref{ap-ryy}) 
by $ j_y(x) = - \partial \mathcal{F} / \partial A_y$.

%
%

%
%
%
%
%
%
%
%
%
%
\end{document}